\documentclass[a4paper]{VisionStyle}
\usepackage{epsfig}

\begin{document}

\title{A cookbook for the XMMSAS}

\author{D.\,Grupe\inst{1}  } 

\institute{
Max-Planck-Institut f\"ur extraterrestrische Physik, Giessenbachstr. D-85748
Garching, Germany}

\maketitle 

\begin{abstract}
This paper will give a short description of a `cookbook' for the XMM-Newton data
reduction software XMMSAS. This 
`cookbook' has been developed at the Max-Planck-Institut f\"ur
extraterrestrische Physik (MPE), Garching. The task of this `cookbook' is to
describe the necessary XMMSAS tasks and show examples to make it easy for new
users of the XMMSAS to understand the steps which
 have to perform to reduce their XMM X-ray data. 

\keywords{Missions: XMM-Newton, macros: XMMSAS }
\end{abstract}

\section{Introduction}
When working with X-ray data each X-ray mission has its own software package to
reduce the data due to the specific requirements of the detectors onboard. For
the XMM-Newton mission the reduction software package XMMSAS has been developed.
The main purpose of the XMMSAS is to allow users to take the Observational Data
File (ODF) and create event files and at a later stage scientific files like
images, spectra and lightcurves. Each XMMSAS task has a manual either in HTML
format or as a postscript file (please see at 
\newline
{\it xmm.vilspa.esa.es/sas/current/doc/packages.All.html} 
\newline
to get the
whole list of available XMMSAS tasks). The manual gives a description of
the tasks and lists all available task parameters. 
However, this is often confusing to new users of the XMMSAS in order 
to see which tasks are needed and how they have to be applied to their data. For
this reason at MPE the idea was born to design a web-based cookbook that
explains users how to work with their XMM-Newton observational data files (ODF) 
in order to create final
scientific files like spectra and light curves. At first this cookbook was only
written for users at MPE, but it has been now designed and make accessible to
users from outside the MPE.

\section{The Cookbook}
\begin{figure*}[ht]
  \begin{center}
    \epsfig{file=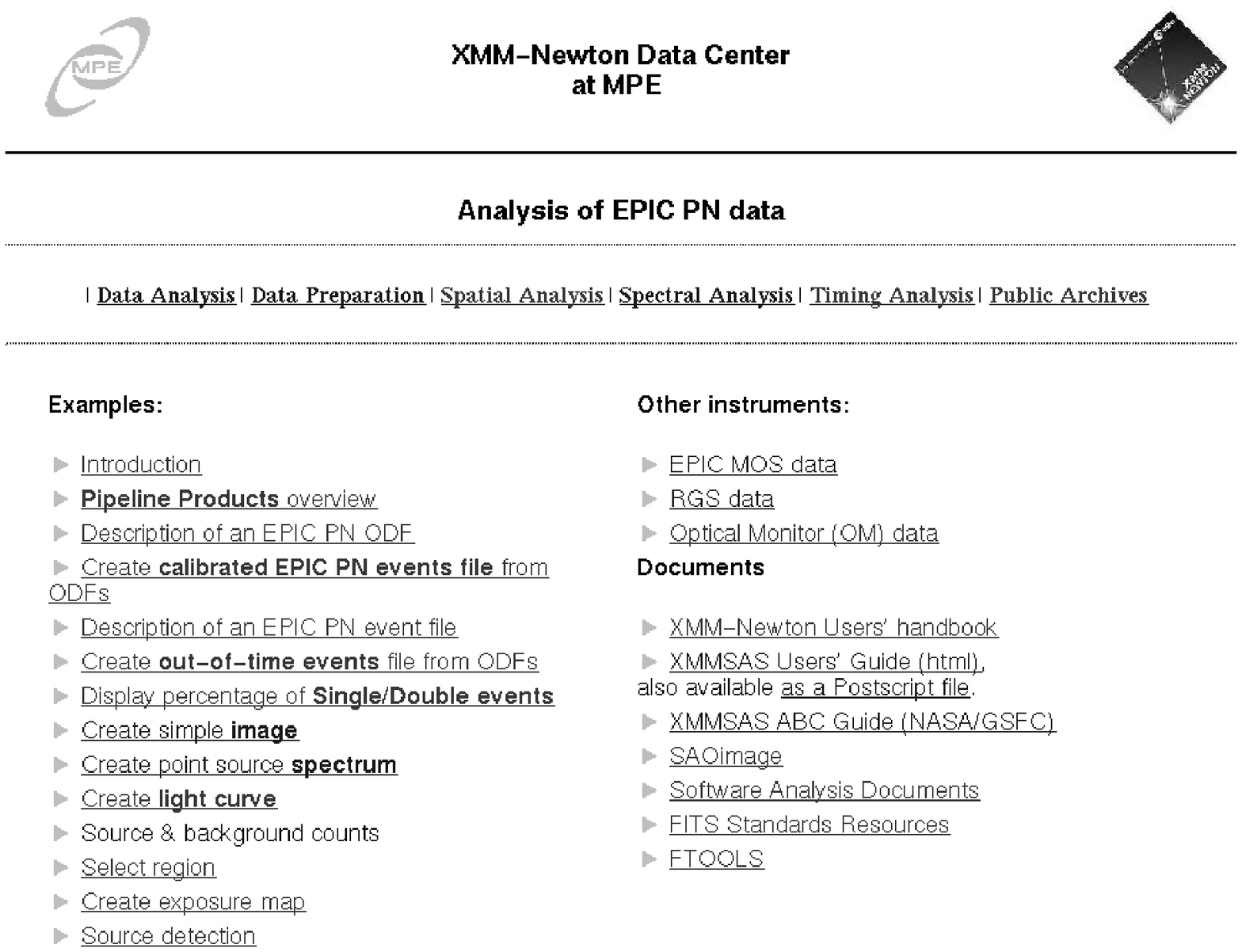, width=17cm}
  \end{center}
\caption{This is an example of a `cookbook' page. This example shows the EPIC PN
page}  
\label{epic_pn_page}
\end{figure*}

To explain the tasks that have to be performed to create event files and finally
scientific files, the cookbook first describes general tasks and is then
organized by the instruments onboard XMM. There are chapters for EPIC PN, EPIC
MOS, the RGS, and the Optical Monitor (OM).The two EPIC chapters have an
introductory part and are then split into sections for spatial, spectral and
timing analysis of the data. Fig \ref{epic_pn_page} displays as an example the
EPIC PN page.
The cookbook has the following structure:

\begin{enumerate}
\item General tasks
\item Data Preparation
\item EPIC PN
\begin{itemize}
\item Spatial analysis
\item Spectral analysis
\item Timing analysis
\end{itemize}
\item EPIC MOS
\begin{itemize}
\item Spatial analysis
\item Spectral analysis
\item Timing analysis
\end{itemize}
\item RGS
\item Optical Monitor
\end{enumerate}

\section{Where to find the Cookbook}
The MPE cookbook for the XMMSAS can be found here:
\newline
{\bf http://wave.xray.mpe.mpg.de/xmm/cookbook/}

\section{Other XMMSAS documents}
The MPE XMMSAS `cookbook' is of course not the only description of the XMMSAS.
First there is ESA's official XMMSAS users guide which is available at VILSPA:
\newline
{\it http://xmm.vilspa.esa.es/sas/documentation}
\newline

Another highly-recommended documentation is the ABC Guide created at NASA/GSFC
which is available in HTML, postscript and pdf format:
\newline
{\it http://heasarc.gsfc.nasa.gov/docs/xmm/abc/abc.html}

\end{document}